\documentclass{eptcs}

\providecommand{\event}{FESCA 2015}

\newlength{\smallskipamounttt}
\usepackage[T1]{fontenc}
\usepackage[latin1]{inputenc}
\usepackage[english]{babel}
\usepackage{amssymb,graphicx,hufflen-macros,ifpdf,listings,rotating,url}

\newcommand{\Dash}{---}
\newcommand{\ffig}{Fig.}
\newcommand{\ffigs}{Figs.}
\newcommand{\surra}[1]{\langle\mathit{#1}\rangle}
\newcommand{\uptoi}[2]{#1_{#2}^{\uparrow}}

\newtheorem{definition}{Definition}
\newtheorem{example}[definition]{Example}
\newtheorem{proposition}[definition]{Proposition}

\title{Using Model-Checking Techniques for Component-Based Systems with
Reconfigurations}
\author{Jean-Michel Hufflen \institute{\logo{femto-st} (\logo{umr}
\logo{cnrs}~6174) \&\ University of Franche-Comté \\ 16, route de Gray; 25030
Besançon Cedex; France} \email{jmhuffle@femto-st.fr}}

\def\titlerunning{Using Model-Checking Techniques for Component-Based Systems
with Reconfigurations}

\def\authorrunning{J.-M. Hufflen}

\begin{document}

\maketitle

\begin{abstract}
Within a component-based approach allowing dynamic reconfigurations, sequences
of successive reconfiguration operations are expressed by means of
reconfiguration paths, possibly infinite. We show that a subclass of such paths
can be modelled by finite state automata. This feature allows us to use
techniques related to model-checking to prove some architectural, event, and
temporal properties related to dynamic reconfiguration. Our method is proved
correct w.r.t.\ these properties' definition.

\end{abstract}

\section{Introduction}

This article aims to show that some properties related to component-based
software with reconfigurations can be proved by implementations based on
model-checking techniques. Let us recall that most of component-based systems
aim to run for a large period of time, so some components may fail or need to
be improved or replaced. That is why dynamic reconfigurations increase the
availability and reliability of such systems by allowing their architecture to
evolve at runtime. Dynamic reconfigurations of software architectures is an
active research topic
\cite{allen-etc1998,bozga-etc2012,krause-etc2011,lanoix-kouchnarenko2014,leger-etc2010},
motivated by practical distributed applications. Such applications are put into
action by means of toolboxes such as \pgname{Fractal} \cite{bruneton-etc2006}.
In addition, if we consider systems with high-safety requirements, the
verification of \emph{architectural}, \emph{event} and \emph{temporal}
properties may be crucial. Some proposals exist, e.g., \cite{falcone-etc2011},
which focus on the verification of properties related to the behaviour of
component-based systems. Within this framework, \cite{dormoy-etc2010} proposes
\logo{ftpl}\footnote{\textbf{F}ractal \textbf{T}emporal \textbf{P}attern
\textbf{L}ogic.}, a temporal logic for dynamic reconfigurations, including such
properties. These properties apply to successive \emph{configurations}\Dash or
\emph{component models}\Dash, chaining reconfigurations being modelled by
\emph{reconfiguration paths}. Since \logo{ftpl} is based on first-order
predicate logic, such properties are undecidable in general, there only exist
partial solutions for proving them.

Within this framework, \cite{kouchnarenko-weber2013,kouchnarenko-weber2014}
developed methods that work whilst software is running and may be reconfigured.
Therefore we know if a property holds step by step, until the current runtime
state. Our method proceeds from a different point of view; our \emph{modus
operandi} is more related to the approach of a procedure's developer when such
a developer aims to prove its procedure before deploying it and putting it into
action. More precisely, we propose a method for verifying such properties, not
at runtime, but on a static abstraction of the reconfiguration model, so we aim
to ensure that such a property holds before the software is deployed and
working, that is, at design-time. In other words, given a reconfiguration path
that may be applied when the software is running, we aim to ensure that a
property holds if this path is actually applied when the software works. Our
method is suitable for finite reconfiguration paths, but also for infinite
ones, provided that they can be modelled by finite expressions, that is, by
using the `\texttt{+}' operator of regular expressions at a final position. In
other words, our method applies to finite reconfiguration paths, and also to
cycles without continuation. This last condition may appear as restrictive, but
in practice, the same sequences are often repeated: a component may be stopped
in some circumstances, restarted in other circumstances, and so on. So the
repetition of identical reconfiguration sequences seems to us to be interesting
in practice, even if they are only a subset of infinite reconfiguration paths.
Besides, our implementation is operational, fits the notion of reconfiguration
path and opens promising ways about properties related to reconfigurations. In
addition, we can prove that this implementation is correct w.r.t.\ the
definitions given in \cite{dormoy-etc2010}. Section~\ref{dkl-reusing} gives
some recalls about the component model we use, our operations of
reconfiguration, and the temporal logic for dynamic reconfigurations. Of
course, most definitions presented in this section come from
\cite{dormoy-etc2010,dormoy-etc2011,dormoy-etc2012,kouchnarenko-weber2013}.
Section~\ref{framework} is devoted to the main outlines of our framework. Then
we give some examples of our programs in Section~\ref{implementation} and study
the correctness of these implementations w.r.t.\ the operators defined in
Section~\ref{dkl-reusing}. In this article, we do not examine the
implementation of all the operators\Dash given in \cite{h2014y}\Dash but our
examples are representative. Section~\ref{discussion} discusses some advantages
and drawbacks of our method, in comparison with other approaches. It also
introduces future work.

\section{Architectural Reconfiguration Model} \label{dkl-reusing}

First we recall how our \emph{component model} is organised. Then we sum up the
operations used for reconfiguring an architecture. Last, we make precise
operators used in \logo{ftpl}, the temporal logic used in
\cite{dormoy-etc2010,dormoy-etc2011,dormoy-etc2012,kouchnarenko-weber2013} for
dynamic reconfigurations.

\subsection{Component Model} \label{component-model}

\begin{figure}[t]

\begin{center}
\ifpdf\includegraphics[width=0.9\linewidth]{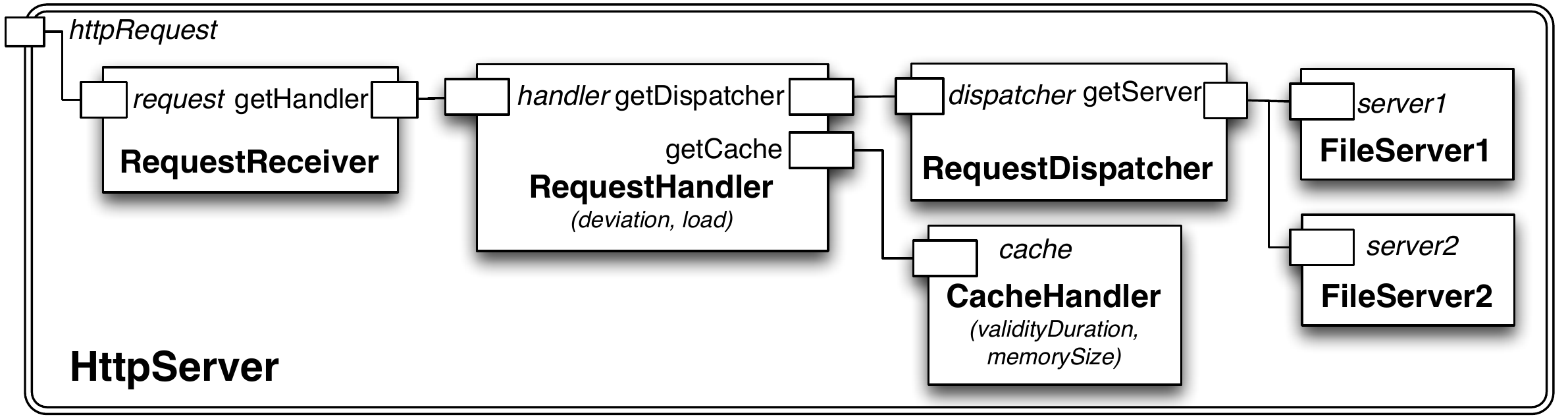}\fi
\end{center}

\caption{Architecture of an \logo{http} server \cite{dormoy-etc2012}.}
\label{hstr-http-server}
\frule
\end{figure}

Roughly speaking, a component model describes an \emph{architecture} of
components. Some simpler components may be subcomponents of a \emph{composite}
one, and components may be \emph{linked}. Let $\mathcal{S}$ be a set of
\emph{class names}\Dash in the sense used in object-oriented programming\Dash a
\emph{component} $\mathcal{C}$ is defined by:
\begin{itemize}
 \item three pairwise-disjoint sets of \emph{parameters}\footnote{Some authors
use the term `attributes' instead. A parameter is related to an internal
feature, e.g., the maximum number of messages a component can process.}
$P_{\mathcal{C}}$, \emph{input port} names $I_{\mathcal{C}}$, and
\emph{output port} names $O_{\mathcal{C}}$;
 \item the class $t_{\mathcal{C}}$ encompassing the services implemented by the
component;
 \item additional functions to get access to the class of a parameter or port
($\tau_{C} : P_{\mathcal{C}} \cup I_{\mathcal{C}} \cup O_{\mathcal{C}}
\rightarrow \mathcal{S}$), or to a parameter's value ($v_{\mathcal{C}} :
P_{\mathcal{C}} \rightarrow \bigcup \mathcal{S}$);
 \item the set $\mathit{sub\text{-}c}_{\mathcal{C}}$ of its subcomponents if 
the $\mathcal{C}$ component is composite. Of course, the binary relation `is a
subcomponent of' must be a direct acyclic graph.
\end{itemize}
A composite component cannot have parameters. The \emph{bindings} of ports form
a set $B$ of couples of output and input port names, being the same type.
Delegation links, between composite component ports and ports of contained
components form a set $D$ of couples, too. As an example of a component-based
architecture, possible components of an \logo{http} server are given in
\ffig~\ref{hstr-http-server}.

\subsection{Configuration Properties}

\begin{example} \label{cache-connected}
Looking at \ffig~\ref{hstr-http-server}'s architecture, we can notice that the
\emph{\texttt{CacheHandler}} component is connected to the
\emph{\texttt{RequestHandler}} component through their respective ports
\emph{\texttt{cache}} and \emph{\texttt{getCache}}. We can express this
\emph{configuration property}\Dash so-called
\emph{\texttt{CacheConnected}}\Dash as follows:
\[B \ni
(\text{\emph{\texttt{cache}}}_{\text{\emph{\texttt{CacheHandler}}}},\text{\emph{\texttt{getCache}}}_{\text{\emph{\texttt{RequestHandler}}}})\]
\end{example}

In fact, such properties\Dash that may be viewed as \emph{constraints}\Dash are
specified using first-order logic formulas over constants (`\texttt{true}',
`\texttt{false}'), variables, sets and functions defined
in~\S~\ref{component-model}, predicates ($=,\in,\ldots$), connectors
($\wedge,\vee,\ldots)$ and quantifiers ($\forall,\exists$). These configuration
properties form a set denoted by $\mathit{CP}$.

\subsection{Reconfiguration Operations} \label{reconfiguration}

\emph{Primitive} reconfiguration operations have been defined, they applied to
a component architecture, and the output is a component architecture,
too\footnote{They may be viewed as \emph{graph transformations} applied to
component models if we consider such models as graphs.}. They are the addition
or removal of a component, the addition or removal of a binding, the update of
a parameter's value. Let us notice that the result of such an operation is
consistent from a point of view related to software architecture: for example,
a component is stopped before it is removed, and removing it causes all of its
bindings to be removed, too. Another point is that these operations are robust
in the sense that they behave like the identity function if the corresponding
operation cannot be performed. For example, if you try to remove a component
not included in an architecture, the original architecture will be returned.
The same if you try to add a component already included in the
architecture\footnote{The reason: the \emph{name} of a component\Dash part of
its definition\Dash can only identify \emph{one} component.}. As a consequence,
these \emph{topological} operations\Dash addition or removal of a component or
a binding\Dash are \emph{idempotent}: applying such an operation twice results
in the same effect than applying it once. General reconfiguration operations on
an architecture are combinations of primitive ones, and form a set denoted by
$\mathcal{R}$. The set of \emph{evolution operations} is
$\mathcal{R}_{\mathit{run}} = \mathcal{R} \cup \{\mathit{run}\}$ where
$\mathit{run}$ is an action modelling that all the stopped components are
restarted and the software is running.

\begin{definition}[\cite{dormoy-etc2011,kouchnarenko-weber2013}]
The operational semantics of component systems with reconfigurations is defined
by the labelled transition system $\mathcal{S} =
\surra{C,C^{0},\mathcal{R}_{\mathit{run}},\rightarrow,l}$ where $C =
\{c,c_{1},c_{2},\ldots\}$ is a set of configurations, $C^{0} \subseteq C$ is a
set of initial configurations, $\mathcal{R}_{\mathit{run}}$ is a finite set of
evolution operations, $\rightarrow\; \subseteq C \times
\mathcal{R}_{\mathit{run}} \times C$ is the reconfiguration relation, and $l :
C \rightarrow \mathit{CP}$ is a total function to label each $c \in C$ with the
largest conjunction of $\mathit{cp} \in \mathit{CP}$ evaluated to `true' over
$\mathcal{R}_{\mathit{run}}$.
\end{definition}

\begin{figure}[t]

\begin{center}
\ifpdf\includegraphics[width=0.9\linewidth]{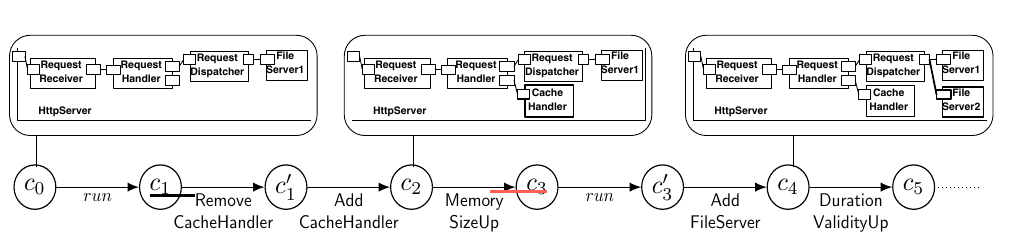}\fi
\end{center}

\caption{Part of an evolution path of \ffig~\protect\ref{hstr-http-server}'s
\logo{http} server architecture \cite{dormoy-etc2012}.} \label{hstr-http-path}
\frule
\end{figure}

Let us note $c \stackrel{\mathit{op}}{\rightarrow} c'$ when a target
configuration $c'$ is reached from a configuration $c$ by an evolution
$\mathit{op} \in \mathcal{R}_{\mathit{run}}$. Given the model $S =
\surra{C,C^{0},\mathcal{R}_{\mathit{run}},\rightarrow,l}$, an evolution path
$\sigma$ of $S$ is a (possibly infinite) sequence of component models
$c_{0},c_{1},c_{2},\ldots$ such that $\forall i \in \mathbb{N}, \exists
\mathit{op} \in \mathcal{R}_{\mathit{run}}, c_{i}
\stackrel{\mathit{op}}{\rightarrow} c_{i + 1} \in\; \rightarrow$. We write
`$\sigma[i]$' to denote the $i$th element of a path $\sigma$. The notation
`$\uptoi{\sigma}{i}$' denotes the suffix path $\sigma[i],\sigma[i + 1],\ldots$
and `$\sigma_{i}^{j}$' ($j \in \mathbb{N}$) denotes the segment path
$\sigma[i],\sigma[i + 1],\ldots,\sigma[j - 1],\sigma[j]$. An example of
reconfiguration path allowing \ffig~\ref{hstr-http-server} to be reached from a
simpler architecture is given in \ffig~\ref{hstr-http-path}
(\ffig~\ref{hstr-http-server}'s architecture is labelled by the $c_{4}$
configuration).

\subsection{Temporal Logic} \label{temporal-logic}

\logo{ftpl} contains events from reconfiguration operations, trace properties,
and temporal properties, respectively denoted by `$\mathit{event}$',
`$\mathit{trace}$', and `$\mathit{temp}$' in the following. Hereafter we only
give some operators of \logo{ftpl}, in particular those used in the
implementations we describe. For more details about this temporal logic, see
\cite{dormoy-etc2011,kouchnarenko-weber2013}. \logo{ftpl}'s syntax is defined
by:
\begin{eqnarray*}
\surra{temp} & ::= & \text{\textbf{after}}\; \surra{event}\; \surra{temp} \mid
\text{\textbf{before}}\; \surra{event}\; \surra{trace} \mid \ldots \\
\surra{trace} & ::= & \text{\textbf{always}}\; \mathit{cp} \mid
\text{\textbf{eventually}}\; \mathit{cp} \mid \ldots \\
\surra{event} & ::= & \mathit{op}\; \text{\textbf{normal}} \mid \mathit{op}\;
\text{\textbf{exceptional}} \mid \mathit{op}\; \text{\textbf{terminates}}
\end{eqnarray*}
where `$\mathit{cp}$' is a configuration property and `$\mathit{op}$' a
reconfiguration operation.
Let $\mathit{cp}$ in $\mathit{CP}$ be a configuration property and $c$ a
configuration, $c$ satisfies $\mathit{cp}$, written `$c \models \mathit{cp}$'
when $l(c) \Rightarrow \mathit{cp}$. Otherwise, we write `$c \not\models
\mathit{cp}$' when $c$ does not satisfy $\mathit{cp}$.

\begin{definition}[\cite{dormoy-etc2011}] \label{evolution-path-df}
Let $\sigma$ be an evolution path, the \logo{ftpl} semantics is defined by
induction on the form of the formulas as follows\footnote{For a complete
definition including all the operators, see \cite{dormoy-etc2011}.}\Dash in the
following, $i \in \mathbb{N}$\Dash:
\end{definition}
\begin{itemize}
 \item for the events:
\begin{center}
\begin{tabular}{l@{\quad if\quad}l}
$\sigma[i] \models \mathit{op}\; \text{\textbf{normal}}$ & $i > 0 \wedge
\sigma[i - 1] \ne \sigma[i] \wedge \sigma[i - 1]
\stackrel{\mathit{op}}{\rightarrow} \sigma[i] \in\; \rightarrow$ \\
$\sigma[i] \models \mathit{op}\; \text{\textbf{exceptional}}$ & $i > 0 \wedge
\sigma[i - 1] = \sigma[i] \wedge \sigma[i - 1]
\stackrel{\mathit{op}}{\rightarrow} \sigma[i] \in\; \rightarrow$ \\
$\sigma[i] \models \mathit{op}\; \text{\textbf{terminates}}$ & $\sigma[i]
\models \mathit{op}\; \text{\textbf{normal}} \vee \sigma[i] \models
\mathit{op}\; \text{\textbf{exceptional}}$
\end{tabular}
\end{center}
 \item for the trace properties:
\begin{center}
\begin{tabular}{l@{\quad if\quad}l}
$\sigma \models \text{\textbf{always}}\; \mathit{cp}$ & $\forall i : i \ge 0
\Rightarrow \sigma[i] \models \mathit{cp}$ \\
$\sigma \models \text{\textbf{eventually}}\; \mathit{cp}$ & $\exists i : i \ge
0 \Rightarrow \sigma[i] \models \mathit{cp}$
\end{tabular}
\end{center}
 \item for the temporal properties:
\begin{center}
\begin{tabular}{l@{\quad if\quad}l}
$\sigma \models \text{\textbf{after}}\; \mathit{event}\; \mathit{temp}$ &
$\forall i : i \ge 0 \wedge \sigma[i] \models \mathit{event} \Rightarrow
\uptoi{\sigma}{i} \models \mathit{temp}$ \\
$\sigma \models \text{\textbf{before}}\; \mathit{event}\; \mathit{trace}$ &
$\forall i : i > 0 \wedge \sigma[i] \models \mathit{event} \Rightarrow
\sigma_{0}^{i - 1} \models \mathit{trace}$
\end{tabular}
\end{center}
\end{itemize}

\begin{example} \label{after-always-example}
If we consider the evolution path of \ffig~\ref{hstr-http-path} again, we can
now express that after calling the \emph{\textsf{AddCacheHandler}}
reconfiguration operation, the \emph{\texttt{CacheHandler}} component is always
connected to the \emph{\texttt{RequestHandler}} component\Dash
\emph{\texttt{CacheConnected}} is the configuration property defined in
Example~\ref{cache-connected}\Dash:
\begin{center}
\emph{\textbf{after \textnormal{\textsf{AddCacheHandler}} normal always
\textnormal{\texttt{CacheConnected}}}}
\end{center}
\end{example}

\section{Our Method's Main Outlines} \label{framework}

\subsection{Basic Idea}

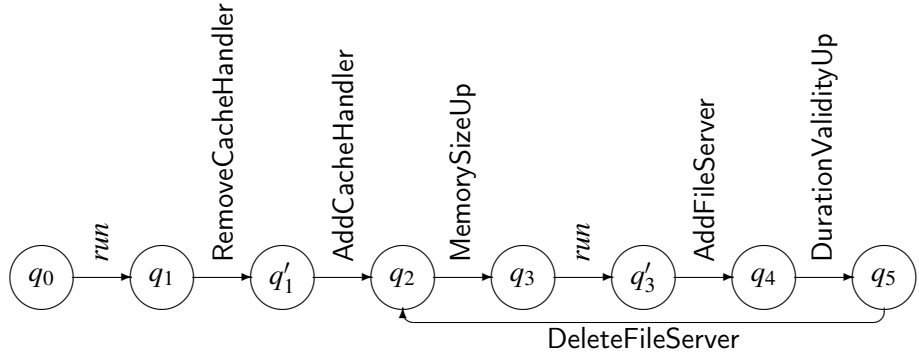
\begin{figure}[t]

\begin{center}
\begin{picture}(30,10)

\put(1,2){\circle{2}}
\put(5,2){\circle{2}}
\put(9,2){\circle{2}}
\put(13,2){\circle{2}}
\put(17,2){\circle{2}}
\put(21,2){\circle{2}}
\put(25,2){\circle{2}}
\put(29,2){\circle{2}}

\put(1,2){\makebox(0,0){$q_{0}$}}
\put(5,2){\makebox(0,0){$q_{1}$}}
\put(9,2){\makebox(0,0){$q'_{1}$}}
\put(13,2){\makebox(0,0){$q_{2}$}}
\put(17,2){\makebox(0,0){$q_{3}$}}
\put(21,2){\makebox(0,0){$q'_{3}$}}
\put(25,2){\makebox(0,0){$q_{4}$}}
\put(29,2){\makebox(0,0){$q_{5}$}}

\put(2,2){\vector(1,0){2}}
\put(6,2){\vector(1,0){2}}
\put(10,2){\vector(1,0){2}}
\put(14,2){\vector(1,0){2}}
\put(18,2){\vector(1,0){2}}
\put(22,2){\vector(1,0){2}}
\put(26,2){\vector(1,0){2}}
\put(21,1){\oval(16,1)[b]}
\put(13,0.9){\vector(0,1){0.1}}

\put(3,2.5){\makebox(0,0)[b]{\begin{turn}{90}
$\mathit{run}$
\end{turn}}}
\put(7,2.5){\makebox(0,0)[b]{\begin{turn}{90}
\textsf{RemoveCacheHandler}
\end{turn}}}
\put(11,2.5){\makebox(0,0)[b]{\begin{turn}{90}
\textsf{AddCacheHandler}
\end{turn}}}
\put(15,2.5){\makebox(0,0)[b]{\begin{turn}{90}
\textsf{MemorySizeUp}
\end{turn}}}
\put(19,2.5){\makebox(0,0)[b]{\begin{turn}{90}
$\mathit{run}$
\end{turn}}}
\put(23,2.5){\makebox(0,0)[b]{\begin{turn}{90}
\textsf{AddFileServer}
\end{turn}}}
\put(27,2.5){\makebox(0,0)[b]{\begin{turn}{90}
\textsf{DurationValidityUp}
\end{turn}}}
\put(21,0){\makebox(0,0){\textsf{DeleteFileServer}}}

\end{picture}

\end{center}

\caption{Reconfiguration path viewed as a finite state automaton.}
\label{hstr-automaton}
\frule
\end{figure}

Our basic idea is that a finite reconfiguration path may be viewed as a
particular case of a \emph{finite state automaton}, more precisely, a kind of
\emph{Büchi automaton}. This property still holds if an infinite
reconfiguration path may be expressed using the `+' operator of regular
expressions at a final position. As an example, let us consider the following
reconfiguration path, related to \ffig~\ref{hstr-http-path}:
\begin{ttfamily}
\begin{tabbing}
$\mathit{run}$ \textnormal{\textsf{RemoveCacheHandler AddCacheHandler}} \\
(\textnormal{\textsf{MemorySizeUp} $\mathit{run}$ \textsf{AddFileServer
DurationValidityUp DeleteFileServer}})+
\end{tabbing}
\end{ttfamily}
it can be modelled by the automaton pictured in \ffig~\ref{hstr-automaton}
(before cycling, the states $q_{0},q_{1},q'_{1},\ldots,q_{5}$ have been
respectively named in connection to the successive component models
$c_{0},c_{1},c'_{1},\ldots,c_{5}$. Let us recall that a finite state automaton
$\mathcal{A}$ is defined by a set $Q$ of \emph{states}, a set $L$ of
\emph{transition labels}, and a set $T \subseteq Q \times L \times Q$ of
\emph{transitions}. Like in Definition~\ref{evolution-path-df} for systems with
reconfigurations, there exists a function $l : Q \rightarrow \mathit{CP}$,
which labels each $q$ state with the largest conjunction of $\mathit{cp} \in
\mathit{CP}$ evaluated to `true' for the $q$ state.

Within this framework, a state of such an automaton modelling a reconfiguration
path is a component model, initial or got by means of successive
reconfiguration operations\Dash primitive or built by chaining primitive
operations\Dash or `\emph{run}' operations. A transition consists of applying
such an evolution operation. Of course, such automata are particular cases:
there is only one initial state, and only one transition can be applied from a
state. More formally, the $T$ set of such an $\mathcal{A}$ automaton satisfies:
\[\forall q,q_{0},q_{1} \in Q, \forall l_{0},l_{1} \in L, (q,l_{0},q_{0}) \in T
\wedge (q,l_{1},q_{1}) \in T \Rightarrow l_{0} = l_{1} \wedge q_{0} = q_{1}\]
In the following, we use the notations `$\mathit{succ}_{\mathcal{A}}(q) =
q_{0}$' or `$q \stackrel{\mathit{succ}_{\mathcal{A}}}{\mapsto} q_{0}$' for the
$q_{0}$ state reached from the $q$ state by a unique transition:
$\mathit{succ}_{\mathcal{A}}(q) = q_{0} \Leftrightarrow \exists! l_{0} \in L,
(q,l_{0},q_{0}) \in T$. If a reconfiguration path is finite, the corresponding
automaton has a final state. Otherwise (like in the example above), there is no
final state in the sense that no transition can be performed. If a state of
such an automaton is reached several times\Dash e.g., the $q_{2}$ state in
\ffig~\ref{hstr-automaton}, reached after $q'_{1}$ and $q_{5}$\Dash considering
that the whole system is back to a previous state is not exact, because some
parameters can have been updated: this is the case in
\ffig~\ref{hstr-automaton}'s example, about the memory's size and duration
validity. As a consequence, some properties related to components' parameters
may not hold. We will go back on this point at the end of~\S~\ref{proofs}.

\subsection{\emph{Modus Operandi}} \label{modus-operandi}

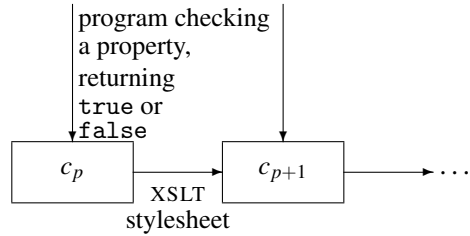
\begin{figure}[t]

\begin{center}
\begin{picture}(15,7.6)

\put(0,1){\framebox(4,2){{\small $c_{p}$}}}
\put(7,1){\framebox(4,2){{\small $c_{p + 1}$}}}
\put(4,2){\vector(1,0){3}}
\put(11,2){\vector(1,0){3}}
\put(2,7.6){\vector(0,-1){4.6}}
\put(9,7.6){\vector(0,-1){4.6}}
\put(2.2,5.3){\makebox(0,0)[l]{{\small \shortstack[l]{program checking \\ a
property, \\ returning \\ \texttt{true} or \\ \texttt{false}}}}}
\put(5.5,1.5){\makebox(0,0)[t]{{\small \shortstack[c]{\logo{xslt} \\
stylesheet}}}}
\put(14.2,2){\makebox(0,0)[l]{\ldots}}

\end{picture}
\end{center}

\caption{Our organisation.} \label{figure-organisation}
\frule
\end{figure}

We use several programming languages within our framework. At first glance,
this choice introduces some complexity, but in reality, each language is used
suitably. \ffig~\ref{figure-organisation} shows how tasks are organised within
our architecture\Dash $(c_{p})_{p \in \mathbb{N}}$ being successive component
models. In our implementation, the \logo{adl}\footnote{\textbf{A}rchitecture
\textbf{D}efinition \textbf{L}anguage.} we use for our component models is
\logo{tacos}+/\logo{xml} \cite{h2013h}. This language using
\logo{xml}\footnote{e\textbf{X}tensible \textbf{M}arkup
\textbf{L}anguage.}-like syntax is comparable with other \logo{adl}s, so from
a theoretical point of view, we could use any \logo{adl}, another example
being \pgname{Fractal}/\logo{adl} \cite{bruneton-etc2006}, but we mention that
the organisation of \logo{tacos}+/\logo{xml} texts make very easy the
programming of primitive reconfiguration operations mentioned
in~\S~\ref{reconfiguration}, that is why we chose this \logo{adl}.
Reconfigurations operations are implemented using
\logo{xslt}\footnote{e\textbf{X}tensible \textbf{S}tylesheet \textbf{L}anguage
\textbf{T}ransformations, the language of transformations used for \logo{xml}
documents \cite{wwwc2007a}. Let us note that if another \logo{adl} is used
within a project, there exist \logo{xslt} programs giving equivalent
descriptions in \logo{tacos}/\logo{xml} \cite{h2013h}. In particular, that is
the case for \pgname{Fractal}/\logo{adl}.}: the input and output are
\logo{tacos}+/\logo{xml} files. When the software is running, only one
component model is in use, so that may be viewed as the identity function
applied to a component model.

A short example of such a \logo{tacos}+/\logo{xml} file is given in
\ffig~\ref{hstr-tacos-xml}. In our implementation, configuration properties are
expressed using \pgname{XQuery} programs \cite{wwwc2008d}, returning `true' or
`false'\footnote{Of course, other choices are possible, in particular
\logo{xslt} stylesheets. Our point of view: \pgname{XQuery} programs are more
concise, what seems to us to be interesting for verifying architectural
properties. Besides, our \pgname{XQuery} programs return `true' or `false', but
we could easily modify them in order to output a counter-example if a property
does not hold.}, as exemplified in \ffig~\ref{hstr-xquery} about the
\texttt{CacheConnected} property. It is well-known that \logo{xml} dialects are
very suitable for specifying architectures\Dash most of \logo{adl}s use this
syntax\Dash and \logo{xslt}/\pgname{XQuery} are very appropriate for operations
modelling reconfigurations and property checks. Chaining reconfigurations and
property checks is expressed by an implementation of automata. There is no
difficulty about the implementation of reconfiguration operations and property
checks, so the descriptions put hereafter concern the part implemented by means
of automata.

\begin{figure}[t]

\begin{lstlisting}
<tacos:components ...> ...
  <tacos:component-specifications>
    <tacos:composite-component id="HttpServer" path="HttpServer">
      <tacos:port ref="Trequest" role="server" name="httpRequest"/> ...
      <tacos:refers-to ref="RequestReceiver"/>
      <tacos:refers-to ref="RequestHandler"/> ...
    </tacos:composite-component>
    <tacos:component id="RequestReceiver" path="HttpServer/RequestReceiver">
      <tacos:port ref="Trequest" role="server" name="request"/> ...
    </tacos:component>
    <tacos:component id="RequestHandler" path="HttpServer/RequestHandler">
      <tacos:port ref="Thandler" role="server" name="handler"/> ...
      <tacos:attributes signature="RequestHandlerAttributes">
        <tacos:attribute name="deviation" value="50"/> ...
      </tacos:attributes> ...
    </tacos:component> ...
  </tacos:component-specifications> ...
  <tacos:binding-specifications>
    <tacos:binding from="request" to="httpRequest" server="RequestReceiver"
                   client="HttpServer"/>
    <tacos:binding from="handler" to="getHandler" server="RequestHandler"
                   client="RequestReceiver"/> ...
  </tacos:binding-specifications>
</tacos:components>
\end{lstlisting}

\caption{Example of a component architecture described by means of
\logo{tacos+/xml}.} \label{hstr-tacos-xml}
\frule
\end{figure}

\subsection{Types Used}

In this paper, we describe our checking functions at a high level. Hereafter we
make precise the types used, in order to ease the reading of our functions. The
formalism we used is close to type definitions in strong typed functional
programming languages like \pgname{Standard ML} \cite{paulson1996} or
\pgname{Haskell} \cite{marlow2010}. Of course, we assume that some types used
hereafter\Dash e.g., `\texttt{bool}', `\texttt{int}'\Dash are predefined.

As abovementioned, an evolution operation is either the identity function,
which expresses that the software is running, or a reconfiguration operation,
which is implemented by applying an \logo{xslt} stylesheet to an \logo{xml}
document and getting the result as another \logo{xml} document. At a
higher-level, such an evolution operation may be viewed as a function which
applies to a component model and returns a component model. Likewise, checking
a property may be viewed as a function which applies to a component model and
returns a boolean value. Assuming that the \texttt{component-model} type has
already been defined, we introduce these two function types as:
\begin{ttfamily}
\begin{center}
\begin{tabular}{l@{ = }l}
\textnormal{\textbf{type}} evolution-operation & component-model $\rightarrow$
component-model \\
\textnormal{\textbf{type}} check-property & component-model $\rightarrow$ bool
\end{tabular}
\end{center}
\end{ttfamily}
Let \texttt{state} be the type used for a state of our automata, starting from
such a state and a configuration\Dash component model\Dash is expressed by the
following type:
\begin{center}
\texttt{\textnormal{\textbf{type}} path-check = state $\times$ component-model
$\rightarrow$ bool}
\end{center}
This \texttt{path-check} type is used within:
\begin{ttfamily}
\begin{center}
\begin{tabular}{l@{ : }l}
\textnormal{\textbf{function}} check-after & evolution-operation $\times$
path-check $\rightarrow$ path-check \\
\textnormal{\textbf{function}} check-always & check-property $\rightarrow$
path-check
\end{tabular}
\end{center}
\end{ttfamily}
In other words, \texttt{check-always($\mathit{check\text{-}p}$)($q$,$c$)}
applies the $\mathit{check\text{-}p}$ function along the $q$ state, its
successor, and so on, starting from the $c$ component model. The result of this
expression is a boolean value. As soon as applying the
$\mathit{check\text{-}p}$ function yields `false', the process stops and the
result is `false'. Likewise,
\texttt{check-after($e$,$\mathit{check\text{-}f}$)($q$,$c$)} also starts from
the $q$ state and the $c$ component model; it applies the
$\mathit{check\text{-}f}$ function as soon as the $e$ event is detected as a
transition of the automata. The property related to the
$\mathit{check\text{-}f}$ function is to be checked for all the component
models resulting from the application of the successive transitions. As a more
complete example, the translation of the formula `\textbf{after} $e$
\textbf{always} $\mathit{cp}$'\Dash where $e$ is an event and $\mathit{cp}$ a
configuration property\Dash is
\texttt{check-after($e$,check-always($\mathit{cp}$))}, which is a function that
applies on a path, starting from a state and component model. The process
starts from the initial state of the automaton. Of course, there are similar
declarations for the functions \texttt{check-before} and
\texttt{check-eventually} (\cf~\S~\ref{temporal-logic}). 

\begin{figure}[t]

\begin{footnotesize}
\begin{ttfamily}
\begin{tabbing}
(: \textnormal{\textsl{Some declarations omitted}. `\texttt{\$filename}'
\textnormal{is the} \logo{xml} \textsl{file analysed.}} :)
\\[\smallskipamounttt]
so\= me \$binding as element(tacos:binding) in \\
 \> doc(\$filename)/tacos:components/tacos:binding-specifications/tacos:binding
\\
 \> sa\= tisfies \\
 \> \> data(\$binding/@from) eq "cache" and data(\$binding/@to) eq "getCache"
and \\
 \> \> data(\$binding/@server) eq "cacheHandler" and data(\$binding/@client) eq
"requestHandler"
\end{tabbing}
\end{ttfamily}

\end{footnotesize}

\caption{The \textbf{CacheConnected} property expressed using \pgname{XQuery}.}
\label{hstr-xquery}
\frule
\end{figure}

\subsection{Ordering States of Automata}

In this section, we introduce some notions related to our automata\Dash they do
not hold about general automata\Dash and used in the following. The states of
our automata modelling reconfiguration paths can be ordered with respect to the
transitions performed before cycling. Let $\mathcal{A}$ be such an automaton,
if $q$ and $q'$ are two states of $\mathcal{A}$:
\[q < q' \stackrel{\text{def}}{\Longleftrightarrow} \exists
(q_{1},\ldots,q_{n}), q \stackrel{\mathtt{succ}_{\mathcal{A}}}{\mapsto} q_{1}
\stackrel{\mathtt{succ}_{\mathcal{A}}}{\mapsto} \cdots
\stackrel{\mathtt{succ}_{\mathcal{A}}}{\mapsto} q_{n}
\stackrel{\mathtt{succ}_{\mathcal{A}}}{\mapsto} q' \text{and
$q,q_{1},\ldots,q_{n},q'$ are pairwise-different.}\]
The notation `$q \le q'$' stands for `$q < q' \vee q = q'$'. The only
transition which does not satisfy this property is the transition going back to
a state already explored, it starts from the state denoted by
$\text{\texttt{q-max}}_{\mathcal{A}}$. If we consider the $\mathcal{A}_{0}$
automaton pictured at \ffig~\ref{hstr-automaton}, $q_{0} < q_{1} < q'_{1} <
q_{2} < q_{3} < q'_{3} < q_{4} < q_{5} =
\text{\texttt{q-max}}_{\mathcal{A_{0}}}$.

\section{Our Method} \label{implementation}

\subsection{Functions}

Our main idea is quite comparable to the \emph{modus operandi} of a
model-checker when it checks the successive states of an automaton in the sense
that we mark all the successive functions of a reconfiguration path. The
possible values of such a mark are:
\begin{description}
 \item[\textnormal{\texttt{unchecked}}] the initial mark for the steps not yet
explored within a reconfiguration path;
 \item[\textnormal{\texttt{again}}] if a universal property (for all the
members of a suffix path) is being checked, it must be checked again at this
step if it is explored again;
 \item[\textnormal{\texttt{checked}}] the property has already been checked,
and no additional check is needed if this step is explored again.
\end{description}
So, such an automaton modelling a reconfiguration path is pre-processed and its
states are marked as \texttt{unchecked}. The mark of a $q$ state is denoted by
\texttt{mark($q$)}. The transition label starting from such a state is denoted
by \texttt{t($q$)}, let us recall that such a transition is the
\texttt{evolution-operation} type, so it can be applied to a component model
$c$ to get the next component model \texttt{t($q$)($c$)}.

We give two implementations of checking properties in
\ffig~\ref{hstr-programs}: the functions \texttt{check-after} and
\texttt{check-always}. We use a high-level functional pseudo-language, except
for updating marks, which is done by means of side effects. A complete
implementation is available at \cite{h2014y}, including other features of
\logo{ftpl}, with similar programming techniques and similar methods for
proving the termination of our functions and the correctness w.r.t.\ the
definitions given in \cite{dormoy-etc2010,dormoy-etc2011}.

\begin{figure}[t]

\begin{footnotesize}
\begin{ttfamily}
\begin{tabbing}
ch\= eck-after($e$,$\mathit{check\text{-}f}$)($q$,$c$) $\longrightarrow$ \\
 \> if mark($q$) == again then true \\
 \> el\= se \qquad // mark($q$) == unchecked \\
 \> \> mark($q$) $\longleftarrow$ again ; $c_{0} \longleftarrow
\mathtt{t}(q)(c)$ ; \\
 \> \> if t($q$) == $e$ then
$\mathit{check\text{-}f}$($\mathtt{succ}_{\mathcal{A}}$($q$),$c_{0}$) else
check-after($e$,$\mathit{check\text{-}f}$)($\mathtt{succ}_{\mathcal{A}}$($q$),$c_{0}$)
\\
 \> \> end if \\
 \> end if \\
end \\[\smallskipamounttt]
check-always($\mathit{check\text{-}p}$)($q$,$c$) $\longrightarrow$ \\
 \> $\mathit{check\text{-}p}(c)\; \wedge$ \= if mark($q$) == checked then true
\\
 \> \> el\= se \qquad // mark($q$) == unchecked $\vee$ mark($q$) == again \\
 \> \> \> mark($q$) $\longleftarrow$ checked ;
check-always($\mathit{check\text{-}p}$)($\mathtt{succ}_{\mathcal{A}}$($q$),t($q$)($c$)) \\
 \> \> end if ; \\
end
\end{tabbing}
\end{ttfamily}

\end{footnotesize}

\caption{Checking properties: two implementations.} \label{hstr-programs}
\frule
\end{figure}

\subsection{Implementations' Correctness} \label{proofs}

\subsubsection{Termination}

\begin{proposition} \label{termination}
The function \emph{\texttt{check-after}} terminates.
\end{proposition}

Let $q_{0}$ be the initial state of our automaton, a principal call of the
\texttt{check-after} function is:
\begin{center}
\texttt{check-after}($e$,$\mathit{check\text{-}f}$)($q_{0}$,$c$)
\end{center}
where $e$ is an event, $\mathit{check\text{-}f}$ a check function being
\texttt{path-check} type, $c$ a component model. Recursive calls of this
function satisfy the invariant:
\[\forall q_{j} : q_{0} \le q_{j} < q_{i}, \mathtt{mark}(q_{j}) =
\mathtt{again}\]
when it is applied to the $q_{i}$ state. Let $q_{k} =
\mathtt{succ}_{\mathcal{A}}(q_{i})$. If $q_{i} < q_{k}$, the invariant holds.
If $q_{i} = \text{\texttt{q-max}}_{\mathcal{A}}$, then $\forall q_{j} : q_{0}
\le q_{j} \le \text{\texttt{q-max}}_{\mathcal{A}}, \mathtt{mark}(q_{j}) =
\mathtt{again}$, that is, the next recursive call applies to a state whose mark
is \texttt{again}. Such a call terminates.

\begin{proposition}
The function \emph{\texttt{check-always}} terminates.
\end{proposition}

This termination proof is similar: a pass is performed by the
\texttt{check-always} function, but this pass may start after the beginning of
a cycle, and the cycle may have to be entered a second time. Globally, two
passes may be needed for an expression such that
\texttt{check-after}$(e,\text{\texttt{check-always}}(cp))$. Before reaching the
end of a cycle, the invariant is:
\[\forall q_{j} : q_{0} \le q_{j} < q_{i}, \mathtt{mark}(q_{j}) =
\mathtt{checked} \vee \mathtt{mark}(q_{j}) = \mathtt{again}\]
when the \texttt{check-always} function is applied to the $q_{i}$ state.
Roughly speaking, when a cycle is performed, the mark has been set either to
\texttt{again}, in which case the property has to be checked again, or to
\texttt{checked}, in which case our function concludes that the temporal
property is true. If the mark has been set to \texttt{again}, it means that the
checking of the temporal property `\textbf{always} $\mathit{cp}$' had not begun
yet; for example, if we were processing the `\textbf{after}' part of
`\textbf{after} $e$ \textbf{always} $\mathit{cp}$'. If re-entering a cycle is
needed, the invariant is:
\[\forall q_{j} :
\mathtt{succ}_{\mathcal{A}}(\text{\texttt{q-max}}_{\mathcal{A}}) \le q_{j} <
q_{i}, \mathtt{mark}(q_{j}) = \mathtt{checked}\]
$q_{i}$ being the current state. Let us recall that
$\mathtt{succ}_{\mathcal{A}}(\text{\texttt{q-max}}_{\mathcal{A}})$ is the first
state of the cycle of the automaton. If the current state reaches the greatest
state w.r.t.\ the '$<$' relation among states, the following recursive call of
\texttt{check-always} is performed with the situation:
\[\forall q_{j} :
\mathtt{succ}_{\mathcal{A}}(\text{\texttt{q-max}}_{\mathcal{A}}) \le q_{j} <
\text{\texttt{q-max}}_{\mathcal{A}}, \mathtt{mark}(q_{j}) = \mathtt{checked}\]
that is, the \texttt{check-always} function terminates at this next call.

\subsubsection{Correctness} \label{correctness}

Concerning the \texttt{check-after} function, let us examine the definition of
the \textbf{after} temporal property: $\sigma \models \mathbf{after}\; e\;
\mathit{temp}\;\; \text{if}\;\; \forall i : (i \ge 0 \wedge \sigma[i] \models e
\Longrightarrow \uptoi{\sigma}{i} \models \mathit{temp})$\Dash where $\sigma$
is a reconfiguration path, $e$ an event, $\mathit{temp}$ a temporal property.
It is sufficient to check this property on the first occurrence of the $e$
event in the transitions handled by our \texttt{check-after} function. The set
$I$ of the $i$ integers such that $i \ge 0 \wedge \sigma[i] \models e$ is a
subset of $\mathbb{N}$. Since $I$ is a subset of a well-ordered set, it has a
smallest element $i_{0}$. So we have $\uptoi{\sigma}{i_{0}} \models
\mathit{temp}$ and $\forall i \in I, i_{0} \le i$. As a consequence,
$\uptoi{\sigma}{i_{0}} \models \mathit{temp} \Longrightarrow \uptoi{\sigma}{i}
\models \mathit{temp}$ for each element $i$ of $I$. Let us consider a call
\texttt{check-after}($e$,$\mathit{check\text{-}f}$)($q_{i}$,$c_{i}$), where
$c_{i}$ is the component model we got in the $q_{i}$ state, and the following
invariant\Dash let us recall that $q_{0}$ is the initial state\Dash:
\[\forall q_{j} : q_{0} \le q_{j} < q_{i}, c_{j} \not\models e\]
holds. If $\mathtt{t}(i) = e$, we check the property implemented by
$\mathit{check\text{-}f}$ from the $q_{i}$ state of the automaton,
corresponding to the suffix of the reconfiguration path starting at the $q_{i}$
state\Dash that is, $\uptoi{\sigma}{i}$ if states are indexed by natural
numbers\Dash which is correct w.r.t.\ the specification. If $\mathtt{t}(i) \ne
e$, the \texttt{check-after} function is recursively called and the invariant
holds. By Proposition~\ref{termination}, if such a call of the
\texttt{check-after} function terminates with \texttt{mark($q_{i}$) = again},
that means that all the automaton's states have been marked \texttt{again}.
Such a mark is put whenever the $e$ event is not found. Consequently, this
event does not appear and the result is the `true' value.

The case of the \texttt{check-always} function is more subtle. Let us recall
that after performing a cycle, the mark is set either to \texttt{again}, in
which case the property has to be checked again, or to \texttt{checked}, in
which case our function concludes that the temporal property is true. This is
not true for any reconfiguration operation, but holds for a large subset. Let
us assume that we have processed a reconfiguration operation $\mathit{op}$ on a
$c$ configuration. If we process $\mathit{op}$ again after cycling throughout
our reconfiguration operations, the current component model may be different
from $c$. In other words, the system is not necessarily in the same state. A
simple counter-example: let us consider \ffig~\ref{hstr-tacos-xml} and a
reconfiguration operation $\mathit{op}_{\text{\texttt{deviation++}}}$ which
increments the \texttt{deviation} attribute of the \texttt{RequestHandler}
component. If this operation is repeated and the property to be
checked is `\textbf{always} \texttt{deviation < 100}', this property will be
true at the first pass but will fail after some iterations. Now let us consider
the reconfiguration operation $\mathit{op}_{\text{\texttt{deviation
$\leftarrow$ 99}}}$ which sets this \texttt{deviation} attribute to a new
value, less than 100. If this reconfiguration operation is repeated, the
property `\texttt{deviation < 100}' will always hold. For our purpose, the
difference between the two operators
$\mathit{op}_{\text{\texttt{deviation++}}}$ and
$\mathit{op}_{\text{\texttt{deviation $\leftarrow$ 99}}}$ is that the latter is
idempotent, not the former. So using a cycle whose global composition of all
the reconfiguration operations is idempotent is sufficient in order for our
\texttt{check-always} function to behave correctly w.r.t.\ the specification of
the `\texttt{always}' operator. It is necessary for checking any property
without further hypotheses. But\Dash as an example\Dash another approach is
accurate if properties to be checked do not deal with parameters' values. The
condition of correctness is the same: the global configuration of all the
reconfiguration operations of the cycle must be idempotent, but reconfiguration
operations dealing with parameters can be ignored. Let us recall that most
reconfiguration operations introduced in~\S~\ref{reconfiguration} are
idempotent. For example, let us recall that if we try to remove a component not
included in a configuration or add a component already included, this
configuration is unchanged. Consequently, applying this operation once or more
causes the same effect. Similar remarks can be done for the addition of a
component, the addition or removal of a binding. In general, the composition of
idempotent operations is not necessarily an idempotent function. In our case,
we can show the composition of \emph{topological} operations is idempotent. To
give a sketch of the proof, let us mention that for two idempotent functions
$f,g : X \rightarrow X$ ($X$ being a set), if $g \circ f = f \circ g$, then $g
\circ f$ is idempotent, too. In fact:
\begin{eqnarray*}
(g \circ f) \circ (g \circ f) & = & g \circ (f \circ g) \circ f \\
                              & = & g \circ g \circ f \circ f \\
                              & = & g \circ f
\end{eqnarray*}
The complete proof is tedious, because many combinations are to be examined.
Some pairs of topological operations yield the identity functions when
composed, other pairs are commutative w.r.t.\ the `$\circ$' composition
operation. Examples coming from \ffig~\ref{hstr-automaton} are the transitions
labelled by the events \textsf{AddFileServer} and \textsf{DeleteFileServer},
which yield the identity function, whereas the transitions labelled by
\textsf{AddFileServer} and \textsf{DurationValidityUp} are commutative.

\section{Discussion and Future Work} \label{discussion}

Within the framework sketched at~\S~\ref{modus-operandi}, our automata have
been implemented using the \pgScheme\ programming language
\cite{sperber-etc2007b}. The complete implementation can be found in
\cite{h2014y}, and \cite{h2014z} describes our functions in a way closer to
this implementation, giving more details about the \pgScheme\ structures we
used. The descriptions of this paper allow us to be more related to a
theoretical model, and to emphasise that our method is close to algorithms
based on marks and used in model-checking, e.g.,
\cite{clarke-etc1986,clarke-etc1994,queille-sifakis1982}. Our implementation
uses \pgScheme\ \emph{streams}\footnote{That is, \emph{potentially} infinite
lists, as implemented within lazy functional programming languages.} for
reconfiguration paths and includes the fact that a reconfiguration path may be
infinite, but only cycles without continuation are processed as shown
in~\S~\ref{implementation}. In practice, our functions are called with an
additional argument for the maximum number of iterations along a
reconfiguration path. If this number is reached, our functions return the part
of the path which is not explored yet and the component model reached, so
end-users can launch the process again if they wish. In this case, we do not
perform complete checking, but only \emph{bounded} checking. With our
implementation, we experienced the property given at
Example~\ref{after-always-example} and the reconfiguration path pictured at
\ffig~\ref{hstr-automaton}; the result is `true', as expected. We also
experienced some variants: for example, if cycling is performed from the
$q_{5}$ state to the $q'_{1}$ state (\cf~\ffig~\ref{hstr-automaton}), the
result is 'true', too, and only the first occurrence of the
\textsf{AddCacheHandler} event is different from the identity function. As a
second variant, if cycling is done from the $q_{5}$ state to the $q_{1}$ state,
the result is `false': when the states $q_{1}$ and $q'_{1}$ are explored by the
\texttt{check-always} function, the property \textsf{CacheConnected} must be
checked and it fails about the $q'_{1}$ state; however, let us notice that such
check after the $q_{2}$ state are not performed if this state is reached after
cycling from the $q_{5}$ state. Practically we have been able to carry out all
the examples of \cite{dormoy-etc2011}; these tests confirmed results expressed
theoretically in that article. Other tests based on reconfigurations of a
location component\footnote{This example is also used in
\cite{h2013h,kouchnarenko-weber2014}.} are successful. From a theoretical point
of view, we explore as few states as possible. In practice, our programs'
efficiency is good, in despite of the communications among several programming
languages. In addition, we plan to study implementation techniques in order to
improve efficiency. We also plan to apply our technique to large-sized case
studies.

The idea of modelling reconfiguration paths by means of automata seems to us to
be very promising, and we plan to go thoroughly into more expressive cases. In
this paper, we are not interested in the \emph{reasons} of reconfigurations:
most often they are implemented by means of \emph{policies} in systems like
\pgname{Fractal}, some examples can be found in \cite{dormoy-kouchnarenko2010}.
Reconfiguration operations may be viewed as operations solving unexpected
situations, but most of these situations are planned by policies and can be
simulated. We plan to enlarge our language of reconfiguration paths, in order
to encompass policies. Likewise, we plan to be able to express
\emph{reconfiguration alternatives}. In other words, we plan to build more
ambitious automata from more expressive reconfiguration paths, provided that we
can check interesting properties. This should lead us to propose new
algorithms, but also to a new version of the temporal operators given
in~\S~\ref{temporal-logic}. The operators given in
\cite{dormoy-etc2011,kouchnarenko-weber2013} refer to a \emph{linear-time}
logic, whereas reconfiguration alternatives should be based on
\emph{branching-time} logic. Explaining this difference is easy:
\cite{dormoy-etc2011,kouchnarenko-weber2013} observe a process in progress, at
runtime, whereas reconfiguration alternative are \emph{possible futures} we
explore before the software is deployed and put into action. Another idea could
be to move the Model Driven Engineering technical space \cite{bezivin2006}, who
would provide more expressive power about model transformations. Other
approaches are closer to a semantic level: for example, \cite{krause-etc2011}
models reconfiguration operations by means of graph rewriting and uses formal
verification techniques along graphs to check properties related to
reconfigurations. A comparable approach is followed in
\cite{kouchnarenko-weber2014}. As another example, \cite{bruni-lavanese2006}
defines a calculus allowing policies. Our approach concerns a more syntactic
level because our reconfiguration operations apply to component models, and
result in other component models. Likewise, we assume that the properties we
check can be verified syntactically on component models: that is the case, at
least for topological properties. Our approach does not cover all cases of
reconfiguration paths, but in practice, the same sequences are often repeated,
as mentioned in the introduction.

\section{Conclusion}

As mentioned above, the starting notions, recalled in~\S~\ref{dkl-reusing},
come from
\cite{dormoy-etc2010,dormoy-etc2011,dormoy-etc2012,kouchnarenko-weber2013}. Our
contribution is our checking method explained in~\S~\ref{framework}.
Practically, it has been actually implemented. Theoretically, we have shown
that our \emph{modus operandi} is correct w.r.t.\ the definitions. Our
contribution also includes the relationship between properties related to
reconfiguration operations and techniques used within model-checking. We think
that using such model-checking techniques for properties related to
component-based software with possible reconfigurations is an open way to
interesting experiments, theoretically as well as practically. Our bridge
between reconfiguration operations and model-checking techniques shows that
tools developed for the analysis of systems can be reused for the analysis of
reconfiguration by simulation of reconfiguration paths. It is well-known that
model-checking techniques can validate a model of a system, not the system
itself. So does our method: as mentioned in the introduction, \emph{we do not
run} components: we only perform a static analysis at design-time. Our approach
cannot replace methods applied at runtime
\cite{kouchnarenko-weber2014,krause-etc2011}, when the software has already
been deployed and is working, but we think that our method can provide some
help at design-time.

\section*{Acknowledgements}

I am grateful to Olga Kouchnarenko and Arnaud Lanoix, who kindly permitted me
to use \ffigs~\ref{hstr-http-server} \&~\ref{hstr-http-path}. Many thanks to
the anonymous referees, who suggested me constructive improvement.

\bibliographystyle{eptcs}
\bibliography{C-et-al,hufflen-2010-,more-xml,new,nfp,other-languages}

\end{document}